\documentclass[sigconf]{acmart}
\setcopyright{none}
\renewcommand\footnotetextcopyrightpermission[1]{} 
\usepackage{fancyhdr}

\begin{document}

\title{Assessing and Addressing Algorithmic Bias -- \\But Before We Get There }
\author{Jean Garcia-Gathright}
\affiliation{%
  \institution{Spotify}
  \city{Somerville} 
  \state{MA} 
}
\email{jean@spotify.com}

\author{Aaron Springer}
\affiliation{%
  \institution{University of California, Santa Cruz}
  \city{Santa Cruz} 
  \state{CA} 
}
\email{alspring@ucsc.edu}

\author{Henriette Cramer}
\affiliation{%
  \institution{Spotify}
  \city{San Francisco} 
  \state{CA} 
}
\email{henriette@spotify.com}

\begin{abstract}
Algorithmic and data bias are gaining attention as a pressing issue in popular press -- and rightly so. However, beyond these calls to action, standard processes and tools for practitioners do not readily exist to assess and address unfair algorithmic and data biases. The literature is relatively scattered and the needed interdisciplinary approach means that very different communities are working on the topic. We here provide a number of challenges encountered in assessing and addressing algorithmic and data bias in practice. We describe an early approach that attempts to translate the literature into processes for (production) teams wanting to assess both intended data and algorithm characteristics and unintended, unfair biases. 
\end{abstract}

\maketitle

\thispagestyle{fancy}
\lhead{\LARGE Adapted from an original publication in the \it{2018 AAAI Spring Symposium Series}}
\rhead{}

\section{Introduction}
Algorithmic fairness and the understanding of its outcomes was anticipated as a research topic as much as twenty years ago ~\cite{friedman1996bias}. Recently, the explosion of widespread machine learning has pushed algorithmic and data bias to the front lines of both the tech press and mainstream media. In parallel, specialized research communities are forming. However, these communities' calls to action are still very hard to apply. Pragmatic methods and tools are necessary to translate nascent research into work in industry practice.

The scattered literature and the proliferation of different communities present industry practitioners with a challenge to keep up, even when highly motivated. We here outline a number of (early) lessons learned from conversations with machine learning-oriented product teams as we think through the pragmatic translation of literature into practice.

\section{Background}
The term bias, in machine learning contexts, is used in somewhat divergent ways. Bias can be framed as unfair discrimination, or as a system having certain characteristics, some intended and some unintended. Any dataset, and any machine learning-based application, is "biased" in the latter interpretation. This means we need to distinguish between intended and unintended/unfair biases. We base our work for practitioners on the pragmatic principle that any dataset is "biased" in some way, that no dataset completely represents the world, and that human decisions in machine learning systems inherently have tradeoffs that can result in (un)intended biases. The goal for product teams is to consider which characteristics of \textit{data}, \textit{algorithms}, and \textit{outcomes} are aligned with the outcomes that they want to achieve. 

We use Olteanu's definition of data bias, "a systemic distortion in the data that compromises its representativeness," as a starting point ~\cite{olteanu2016social}. Social data encompasses content generated by users, relationships between those users, and application logs of user behaviors. Olteanu's framework comprehensively examines biases introduced at different levels of social data gathering and usage, including: user biases, societal biases, data processing biases, analysis biases, and biased interpretation of results. 

Other taxonomies of algorithmic and data bias highlight the interplay between data bias and algorithmic bias: biased training data results in biased algorithms, which in turn produce more biased data in a feedback loop. The Baeza-Yates taxonomy consists of 6 types of bias: activity bias, data bias, sampling bias, algorithm bias, interface bias, and self-selection bias ~\cite{baeza2016data}. These biases form a directed cycle graph; each step feeds biased data into the next stage where additional characteristics are introduced. This cyclical nature makes it difficult to discern where to intervene; models like Baeza-Yates' help break down the cycle and find likely targets for initial intervention.

As a definition for outcome bias, we use the description of "computational bias" from Friedman and Nissenbaum: "Discrimination that is systemic and unfair in favoring certain individuals or groups over others in a computer system" ~\cite{friedman1996bias}. 

\section{Translation into Bias Identification Processes}
A major challenge is translating the growing, but scattered literature into a step-by-step process that works in practice. The first step to correcting algorithmic biases is identification of potential biases, for which we have three possible entry points:
\begin{itemize}
\item Biases in input data
\item Computational biases that may result from algorithm and team decisions
\item Outcome biases, for example for specific user groups
\end{itemize}

Teams need methods to help them ask concrete questions about algorithmic and data biases in their product. In our case, we used the existing bias frameworks above, and translated them into a easier to digest summary checklist of characteristics to consider for data, models, and outcomes. Each row in the checklist describes a bias category; the checklist user then assesses: how the bias may affect the project's outcome, how to prioritize mitigation of the bias, and potential solutions.

However, this is not enough; looking for candidate biases will surface a large number, such that it becomes difficult to determine which issues to tackle first and which are better suited as long term goals. Furthermore, biases may compound and interact. It is essential that these bias targets are prioritized by evaluating the impact for different stakeholders and anticipating future compounding effects. Weighing these bias targets against each other involves a complex decision involving level of harm, ubiquity of bias, and business driven priorities. After assessment, very specific domain and organizational knowledge will be necessary to deliver concrete methods and recommendations. 

\section{Domain challenges: case study in voice}
In addition to domain-agnostic frameworks and tools for understanding algorithmic bias in general, domain-specific investigations may also be necessary in practice. For example, voice interfaces are rapidly gaining popularity, but, unfortunately, voice interfaces may amplify bias due to their unique affordances. For example, voice interfaces may struggle with regional accents ~\cite{best2013recognizing}. Language dialects also may result in worse accuracy and voice recognition ~\cite{tatman2017gender}. Even if dialects and accents were perfectly recognized by voice interfaces, these interfaces would still struggle to leverage common solutions from other modalities. For example, recommender systems often suffer from popularity bias. Solutions to enhancing discoverability of the long tail of content include increasing serendipity and novelty among recommendations ~\cite{vargas2011rank}. Unfortunately, users of voice interfaces are often trying to accomplish a task quickly, and listing ten search results that include some popular, some novel, and some serendipitous results may degrade the user experience because of the time it takes to verbally list them. Therefore, this task of countering popularity bias may be much harder in voice where only one result is returned.

A major struggle with many types of bias research is understanding whether the metric differences measured are due to algorithmic/data bias or simply due to natural demographic variation ~\cite{mehrotra2017auditing}. One way to measure and correct bias in this case may be finding problems where a ground truth answer is available. Springer et al. examine the types of content that current voice interfaces underserve due to content characteristics ~\cite{springer2018play}. For example, current voice interfaces often transcribe dialect speech into Standard American English; this can result in a user asking for a music track titled "You Da Baddest" and the voice interface transcribing and searching for "You're the baddest" which may not result in finding the intended track. These entity resolution difficulties fortunately mean that some form of ground truth may sometimes be available. In this case, we can tease apart the algorithmic bias from demographic differences and identify ways to correct issues. However, there is no ground truth of human experience, nor behavior.

\section{Pragmatic Challenges}
In this section, we present a few examples of pragmatic challenges that may be encountered when attempting to mitigate data and algorithmic bias in an industry setting.

\subsection{Prioritizing Correcting Bias}
Engineering teams abide by a carefully planned roadmap of deliverables, with much energy devoted to maintaining their current systems and pushing new features to product. Setting aside time to measure and correct bias has to compete with other pressing priorities. Furthermore, in a situation where features built from imperfect data have already been surfaced in the product, significant changes in the feature may be perceived as too risky. Framing such work in terms of business goals, such as improving performance across markets and improvement of quality, is a compelling argument for pursuing this work (compared with, for example, unspecified appeals that addressing bias should be important). 

\subsection{Proposing Minimum Viable Products}
Agile development is a popular approach to product development. In an Agile-style environment, there is an emphasis on quick delivery of minimum viable products followed by continuous iteration. In order to translate research on bias to solutions in product, it is necessary to propose a minimal solution that can be delivered and then improved. For example, is it possible to move forward with solutions on narrow use cases or with imperfect measurements? Caution is required here, to prevent the minimum viable product from simply being accepted as the final product. Long-view thinking is also necessary, so that even as imperfect products are delivered quickly, there is still a path of iteration toward a more ideal solution. 

\subsection{Addressing Technical Debt via Cultural Changes}
In the early stages of a company's development, the issue of scaling globally seems impossibly distant. In this scenario, teams may accumulate technical debt as a result of limited access to resources and data. When company growth reaches a point where global scaling becomes a priority, new perspectives and attitudes are necessary. Diversity in hiring becomes even more important. Longer-term cultural change and education toward bias-awareness would also encourage engineers to design models and features with delivery to a global audience in mind, avoiding bias-related technical debt at the outset of the design process.

\section{Discussion}
To assess and address algorithmic biases, we need to translate the growing literature into methods that are applicable across domains and easy to communicate. Teams need lightweight tools to make these processes their own, rather than responding to calls to action from elsewhere. We have described our early attempt to translate literature on algorithmic bias, and identified the domain-specific and pragmatic challenges that follow. Shared understanding within industries and researcher communities, including the sharing of developed methods and lessons learned, combined with a bottom-up application of understandable frameworks by teams themselves, appears most fruitful.

\bibliographystyle{abbrv}
\bibliography{fatrec_2018.bib}

\end{document}